\documentclass[preprint,11pt, authoryear]{elsarticle}

\usepackage{setspace}
\usepackage{natbib}
\usepackage{amsthm}
\usepackage{array}
\newtheorem{theorem}{Theorem}

\usepackage{booktabs}
\usepackage{multirow}
\usepackage[T1]{fontenc}
\usepackage{amssymb}
\usepackage{appendix}
\usepackage{amsmath}
\usepackage{longtable}
\usepackage[utf8]{inputenc}
\usepackage[T1]{fontenc}
\usepackage{xcolor}
\usepackage{amsmath,amssymb,amsfonts}
\usepackage{textcomp}
\usepackage{float}
\usepackage{array}
\usepackage{booktabs} 
\usepackage{bm}
\usepackage{epsfig}
\usepackage{soul}
\usepackage{pgf}
\usepackage{multirow}
\usepackage{bigstrut}
\usepackage{comment}
\usepackage{rotating}
\usepackage{pdflscape}
\usepackage{amsmath} 
\usepackage{lscape}
\usepackage{subcaption}
\usepackage{setspace}
\usepackage{amssymb}
\usepackage{amsmath}
\usepackage{relsize}
\usepackage{url}
\usepackage{algorithm}
\usepackage{algpseudocode}
\usepackage{natbib}
\usepackage{graphicx}
\usepackage{multirow}
\usepackage{multicol}
\usepackage{comment}
\usepackage{booktabs} 
\usepackage{array} 
\usepackage{balance} 
\usepackage{soul}
\usepackage{siunitx}
\usepackage{graphicx}
\usepackage{subcaption} 
\usepackage{float}
\usepackage{xcolor}
\usepackage{comment}
\usepackage{graphicx}
\usepackage{tikz}
\usetikzlibrary{positioning, shapes, arrows.meta}

\usetikzlibrary{positioning,fit}

\usepackage{pifont}

\newcommand{\cmark}{\textcolor{green!60!black}{\ding{51}}}
\newcommand{\xmark}{\textcolor{red}{\ding{55}}}

\newcolumntype{C}[1]{>{\small\centering\arraybackslash}m{#1}}
\newcolumntype{L}[1]{>{\small\raggedright\arraybackslash}m{#1}}

\usepackage{array}

\usepackage[dvipsnames]{xcolor}
\usepackage{pifont}
\usepackage{xurl}
\usepackage[margin=2.5cm]{geometry}
\allowdisplaybreaks
\usepackage[final]{hyperref}
\makeatletter
\providecommand*{\toclevel@subsubsubsection}{4}
\providecommand*{\l@subsubsubsection}{\@dottedtocline{4}{7em}{4em}}
\makeatother
\usetikzlibrary{shapes.geometric}
\usepackage{titlesec}

\titleclass{\subsubsubsection}{straight}[\subsubsection]

\newcounter{subsubsubsection}[subsubsection]
\renewcommand\thesubsubsubsection{\thesubsubsection.\arabic{subsubsubsection}}

\titleformat{\subsubsubsection}
  {\normalfont\normalsize\bfseries}
  {\thesubsubsubsection}{1em}{}

\titlespacing*{\subsubsubsection}
  {0pt}{3.25ex plus 1ex minus .2ex}{1.5ex plus .2ex}

\makeatletter
\providecommand*{\toclevel@subsubsubsection}{4}
\makeatother

\journal{a journal}

\definecolor{myblue}{HTML}{000B4F}
\definecolor{myblue2}{HTML}{20368F}
\definecolor{myblue3}{HTML}{829CD0}
\definecolor{myblue4}{HTML}{C9D6F0} 
\definecolor{myblue5}{HTML}{E6ECFA} 
\definecolor{myborder}{HTML}{1C2A5A}

\begin{document}

\begin{frontmatter}

\title{
Cutting-plane methodology via quantum optimization for solving the Traveling Salesman Problem\\
}

\author[aff1]{Alessia Ciacco}
\ead{alessia.ciacco@unical.it}

\author[aff2]{Luigi Di Puglia Pugliese}
\ead{luigi.dipugliapugliese@icar.cnr.it}

\author[aff1]{Francesca Guerriero}
\ead{francesca.guerriero@unical.it}

\affiliation[aff1]{organization={Department of Mechanical, Energy and Management Engineering, University of Calabria},
           city={Rende},
            postcode={87036},
            state={CS},
            country={Italy}}

\affiliation[aff2]{organization={Istituto di Calcolo e Reti ad Alte Prestazioni, Consiglio Nazionale delle Ricerche (ICAR-CNR)},
           city={Rende},
           postcode={87036},
           state={CS},
           country={Italy}}

\begin{abstract}
The Traveling Salesman Problem is a classical NP-hard combinatorial optimization problem that has been extensively studied in operations research. A major challenge in Traveling Salesman Problem formulations is the large number of subtour elimination constraints required to ensure a valid tour. To address this issue, we adopt an iterative approach grounded in well-established operations research techniques, in which subtour elimination constraints are generated dynamically. In addition, we integrate a preprocessing phase to reduce the number of candidate arcs. In this work, we investigate both classical and quantum optimization approaches for solving the problem using the proposed framework. In particular, for quantum optimization we analyze quantum annealing techniques within the D-Wave framework, considering both direct quantum execution on the QPU and hybrid quantum–classical solvers. Computational experiments show that the proposed strategies significantly reduce the model size and lead to positive improvements in computational performance across classical, direct quantum, and hybrid optimization approaches.
\end{abstract}


\end{frontmatter}
\section{Introduction}
The Traveling Salesman Problem (TSP) is a fundamental problem in combinatorial optimization, that has been widely studied. The problem consists of determining a minimum-length tour that allows a traveler to visit a set of cities exactly once and finally return to the starting city. Despite its extremely simple formulation, the TSP represents a fundamental paradigm for a wide class of graph optimization problems and finds applications in numerous real-world contexts. Moreover, many variants of complex industrial problems can be reduced to formulations analogous to the TSP or to its extensions.
One of the most common formulations of the TSP is based on Integer Linear Programming (ILP). In these formulations, the decision variables typically represent the selection of arcs in the graph that form the tour. In addition to the constraints imposing that each city is visited exactly once, it is necessary to introduce additional constraints known as subtour elimination constraints. These constraints ensure that the final solution consists of a single cycle visiting all cities and prevent the formation of smaller cycles involving only subsets of nodes. However, the potential number of such constraints grows exponentially with the number of cities. Consequently, explicitly including all subtour elimination constraints in a complete formulation becomes impractical even for moderately sized instances. Efficiently handling these constraints therefore represents one of the most critical and challenging aspects in modeling and solving the TSP.
This inherent difficulty is a direct reflection of the problem's theoretical complexity. From a computational complexity perspective, the problem belongs to the class of NP-hard problems \cite{karp1972reducibility}, meaning that no polynomial-time algorithm is known for the general case. As the number of cities increases, the number of feasible tours grows factorially. Such a combinatorial explosion explains why the TSP continues to attract significant research attention. Over the years, numerous approaches have been proposed to address the problem, ranging from exact algorithms to heuristic and metaheuristic methods. 
In recent years, the increasing availability of new computational architectures has further opened the possibility of exploring alternative paradigms for solving complex optimization problems.

In this context, quantum computing represents a particularly promising research direction for tackling complex combinatorial optimization problems. More broadly, the field of quantum optimization investigates how quantum mechanical phenomena, such as superposition and tunneling, can be exploited to explore large solution spaces more efficiently than classical algorithms. For a comprehensive overview of existing applications of quantum optimization approaches, we refer to the survey by \cite{ciacco2025review}, which provides a broad review of quantum algorithms across multiple application domains, including healthcare, finance, production planning, and logistics.
Driven by these diverse requirements, the development of quantum optimization has predominantly followed two main research directions. The first is based on gate-model quantum computing, which employs quantum algorithms designed for universal quantum processors. In this paradigm, computation is performed through a sequence of discrete quantum gates applied to a set of qubits. A prominent class of approaches within this framework is represented by Variational Quantum Algorithms (VQAs), which combine quantum circuits with classical optimization routines. This class includes algorithms such as the Quantum Approximate Optimization Algorithm (QAOA) and the Variational Quantum Eigensolver. 
The second relies on Quantum Annealing (QA), a paradigm that exploits quantum effects, such as tunneling, to identify minimum-energy configurations corresponding to optimal or near-optimal solutions. QA has therefore been explored in a wide range of combinatorial optimization problems, including scheduling~\citep{venturelli2016job, perez2024solving, ciacco2025Educational}, routing~\citep{ciacco2025steiner, holliday2025advanced, osaba2025quantum}, facility location~\citep{ciacco2026facility, malviya2023logistics}, and packing~\citep{de2022hybrid, garcia2022comparative}.
In recent years, dedicated hardware systems for QA have been developed and made accessible, including those produced by D-Wave Systems. To exploit these architectures, problems must be formulated in representations compatible with the computational model of the quantum hardware. In particular, the ecosystem developed by D-Wave supports several quadratic optimization models, including the Quadratic Unconstrained Binary Optimization (QUBO) model and the Constrained Quadratic Model (CQM), which represent two different ways of expressing combinatorial optimization problems in a form compatible with QA.

In the QUBO model, the problem is formulated as the minimization of a quadratic function defined over binary variables. In this representation there are no explicit constraints: all problem conditions must be incorporated into the objective function through appropriate penalty terms. This approach is particularly suitable for direct execution on the D-Wave quantum processing units (QPUs), since the hardware is designed to minimize quadratic functions over binary variables represented as quantum spins. In practice, the QUBO problem is translated into an Ising model and then mapped onto the physical architecture of the QPU through embedding procedures based on the hardware connectivity graph. However, this transformation may introduce several difficulties. Additional limitations arise from both the penalty-based encoding of constraints and the restricted connectivity of current QPUs. The former can make the objective function highly sensitive to the choice of penalty coefficients, while the latter often requires chains of logical qubits, increasing the effective problem size and potentially degrading solution quality.

Furthermore, the D-Wave ecosystem also provides hybrid solvers that combine classical optimization techniques with the use of the quantum annealer. 
This strategy makes it possible to handle significantly larger problems than those that could be executed directly on the QPU alone, using the QPU as a component within a broader optimization process. Within this context, the CQM represents a more recent modeling formalism introduced within the D-Wave ecosystem. It allows linear or quadratic constraints to be represented explicitly and separately from the objective function. Unlike the QUBO model, constraints do not necessarily need to be transformed into penalty terms, enabling a more natural and often more compact problem formulation. CQM models are generally solved using hybrid solvers that combine classical optimization algorithms with techniques inspired by QA. This approach allows problems with a larger number of variables and constraints to be addressed compared to what would be feasible on a pure QPU, while still enabling the exploitation of quantum components during the solution search process. In practice, these hybrid solvers are typically accessed as black-box optimization tools, where the internal orchestration between classical and quantum resources is managed automatically by the D-Wave framework.

Thus, QUBO and CQM represent two complementary modeling approaches within the D-Wave ecosystem. The QUBO model is more directly compatible with execution on the QPU and requires the transformation of constraints into penalties within the objective function, whereas the CQM model allows the explicit representation of constraints and is typically solved using hybrid solvers that integrate classical and quantum computational resources. The choice between these two approaches therefore depends both on the nature of the problem and on the desired computational characteristics.

However, despite this potential, several practical and theoretical limitations still hinder the effective application of current quantum optimization approaches. In particular, the associated energy landscapes often exhibit high ruggedness and numerous local minima, which can significantly affect the convergence of both QA and variational algorithms~\citep{abbas2024challenges}. Moreover, recent analyses have highlighted the gap between demonstrating the feasibility of quantum formulations and achieving any meaningful notion of quantum advantage. In particular, \citep{smith2025travelling} argue that current quantum optimization pipelines remain highly sensitive to modeling choices, penalty parameter tuning, hardware noise, and embedding constraints. As a result, these approaches currently show no clear scaling benefits over state-of-the-art classical methods.

In this work, we investigate both classical and quantum optimization approaches for solving the TSP. Classical formulations are solved using a commercial optimization solver, while quantum models are implemented within the QA framework using both direct execution on the QPU and hybrid quantum–classical solution approaches based on QUBO and CQM formulations.
The presence of subtour elimination constraints, however, poses a significant challenge in both cases, since the number of such constraints is extremely large. To address this, we adopt a lazy constraint-generation approach, a well-established technique in combinatorial optimization literature for handling constraints that are too numerous to be included a priori {\cite{dantzig1954solution}}. Following this iterative paradigm, the initial model is built with only the set of degree constraints. The solutions generated by the solver are then analyzed to detect possible subtours present in the resulting tour. Whenever subtours are identified, the corresponding constraints are added to the model, and the problem is solved again. This iterative process is repeated until a solution without subtours is obtained. { This method is known as a ``cutting-plane'' approach because it works by solving a relaxed version of the problem and then iteratively adding constraints, the ``cuts'', that remove the current infeasible solution without eliminating any valid Hamiltonian tours.} This approach significantly reduces the size of the initial model and introduces only the constraints that are actually required during the optimization process. To the best of our knowledge, this work represents the first application of an iterative lazy constraint generation approach within a QA framework to improve the performance and scalability of the TSP. Furthermore, in order to further improve the efficiency of the model and reduce the size of the problem to be solved, we integrate in our approach a preprocessing phase based on the Cost-based Arc Filtering (CAF) method proposed in~\cite{ciacco2026CAF}. This method was developed with the aim of reducing the solution space of the TSP by preliminarily eliminating arcs that cannot belong to an optimal tour. The core idea of this approach is to analyze the structure of the graph and the distances between cities in order to identify arcs that are dominated or incompatible with an optimal solution. 
The integration of the CAF preprocessing method with the iterative generation of subtour elimination constraints therefore combines two complementary strategies for reducing complexity: on the one hand the reduction of candidate arcs in the graph, and on the other the dynamic generation of only those subtour constraints that are actually required. This combination contributes to making the model more compact and better suited to the characteristics of current QA architectures.
The substantial reduction in problem complexity achieved through these techniques allows for bridging the gap between theoretical quantum optimization and practical applications. Specifically, this approach enables the resolution of standard TSP benchmark instances within a QA framework, marking a first in the literature for such a class of problems.

The paper is organized as follows. Section~\ref{sec:Related works} reviews the main contributions in the literature on the TSP and its solution approaches, with a particular focus on quantum-based optimization techniques. Section~\ref{sec:problem_statement} provides the formal description of the TSP and presents its mathematical formulation, including a discussion on the complexity and the role of subtour elimination constraints.
Section~\ref{sec:solution_approach} describes the proposed solution framework, including the iterative approach for handling subtour constraints and a brief overview of the CAF preprocessing method adopted. 
Section~\ref{sec:experiments} reports the computational experiments, including the solver setup and the numerical analysis performed using both classical and quantum approaches. 
Finally, Section~\ref{sec:conclusions} summarizes the main findings of this work and outlines possible directions for future research.

\section{Related works}\label{sec:Related works}
This section provides a brief overview of the state of the art in order to position the proposed research within the broader scientific context. The literature is reviewed along two main directions: classical approaches for the TSP and recent studies on QA methods for routing and graph-based combinatorial problems. 

\subsection{Classical Approaches}
One of the earliest fundamental contributions to the study of the TSP is the work of ~\cite{dantzig1954solution}, which introduced a linear programming approach based on cutting planes for solving instances of the problem. 
Subsequently, \cite{miller1960integer} proposed an integer programming formulation, {called} the MTZ formulation, which introduces auxiliary variables to prevent the formation of subtours. 
\cite{laporte1992tsp} provides a thorough survey of both exact and approximate approaches. A comprehensive computational analysis of exact methods and their implementation is presented, also, by \cite{applegate2006traveling}, who describe the development of advanced solvers and the algorithmic strategies used to address large-scale TSP instances. From a broader perspective, \cite{cook2010integer} traces the evolution of combinatorial integer programming, highlighting the pivotal role of the TSP in the advancement of the field.
Given the computational complexity of the problem, many studies have focused on {heuristic methods that produce} high-quality solutions {in reasonable time}. 
\cite{croes1958tsp} proposed a local improvement procedure based on edge exchanges, known as the 2-opt heuristic. 
\cite{karg1964heuristic} introduced additional heuristic approaches based on constructive strategies and local search techniques. 
\cite{lin1973effective} employ an adaptive sequence of k-opt exchanges and remains one of the most effective heuristics for solving large TSP instances.

\subsection{Quantum Approaches}\label{sec:quantum}
Some studies have investigated the application of quantum computing techniques to the TSP, typically focusing on small problem instances due to the limitations of current quantum hardware. A key aspect of these works concerns the size of the considered instances 
and the mechanism adopted to prevent the formation of subtours.

{VQAs} have been explored for very small TSP instances. For example, ~\cite{qian2023qaoatsp} investigate QAOA-based formulations for symmetric TSP problems using different mixer Hamiltonians. 
In their formulation, the tour is represented using binary variables indicating the position of each city, while constraints are {encoded in the Hamiltonian via penalty terms}. Subtour elimination is therefore handled indirectly {through these penalties}.
Similarly, ~\cite{bell2025qaoarepresentations} analyze several Hamiltonian encodings of routing problems within QAOA {, comparing} quadratic QUBO formulations and higher-order Hamiltonians. Quadratic encodings are {hardware-compatible but require many qubits and penalty terms, while higher-order encodings reduce variables at the cost of increased circuit complexity.} In both cases, { subtours are handled by adding penalty terms to the Hamiltonian (the function that assigns a cost to each candidate route), so that invalid solutions receive a higher energy. }
~\cite{baldazzi2025variationaltsp} propose a variational algorithm based on entangled quantum registers encoding city-to-city connections and demonstrate the approach on a four-city TSP instance implemented on a silicon photonic quantum processor. 
The imposed correlations reduce the space of infeasible solutions, mitigating subtour formation.

One of the earliest implementations of the TSP on a quantum annealer is presented by ~\cite{jain2021tsp}, where the problem is formulated as a QUBO model and executed on a D-Wave processor. Due to hardware connectivity constraints and embedding requirements, the experiments are limited to instances involving approximately eight cities. 
~\cite{warren2020tsp} evaluates software approaches for solving the TSP on D-Wave quantum annealers{, showing} that realistic instances cannot be handled directly by current hardware. 
Moreover, solutions produced by the annealer may contain multiple disconnected cycles, highlighting the difficulty of encoding subtour elimination constraints within QUBO formulations. 
~\cite{bochkarev2026quantum} further analyze the scalability of QUBO models{, demostrating} 
that encoding global constraints like subtour elimination significantly increases the number of binary variables, leading to large embeddings that exceed the capacity of current quantum annealers. 
~\cite{ciacco2026CAF} propose a preprocessing strategy aimed at reducing the number of decision variables in the optimization model while preserving the existence of Hamiltonian cycles. The resulting formulation is solved using both classical optimization methods and a hybrid quantum–classical approach based on the CQM framework. 
{Results} 
show that the preprocessing step reduces the model size and improves performance.

Overall, the literature indicates that current quantum approaches to the TSP remain restricted to very small instances. 
The main limitations arise from qubit availability 
and the difficulty of encoding complex combinatorial constraints such as subtour elimination in energy-based formulations. {To provide a comparison, Table~\ref{TabellaQuantum} summarizes the main quantum approaches proposed in the literature, highlighting their methodological differences, subtour constraint handling, and scalability.}

{\small
\begin{table}[ht]
\centering
\renewcommand{\arraystretch}{1.25}
\begin{tabular}{
    L{4.2cm}
    C{1.2cm}
    C{1.2cm}
    C{1.2cm}
    C{1.8cm}
    C{2.2cm}
    C{2.0cm}
}
\hline
\textbf{Reference} &
\textbf{VQA} &
\multicolumn{2}{c}{\textbf{QA}} &
\textbf{SEC} &
\textbf{Benchmark instances} &
\textbf{Max size} \\

\cline{3-4}
&
&
\textbf{Direct QPU} & \textbf{Hybrid} &
& & \\
\hline

~\cite{qian2023qaoatsp} &
\cmark & \xmark & \xmark &
\cmark &
\xmark &
3--5 nodes \\

~\cite{bell2025qaoarepresentations} &
\cmark & \xmark & \xmark &
\xmark &
\xmark &
$\leq$ 5 nodes \\

~\cite{baldazzi2025variationaltsp} &
\cmark & \xmark & \xmark &
\cmark &
\xmark &
4 nodes \\

~\cite{jain2021tsp} &
\xmark & \cmark & \xmark &
\xmark &
\xmark &
8 nodes \\

~\cite{warren2020tsp} &
\xmark & \cmark & \xmark &
\xmark &
\xmark &
$\leq$ 8 nodes \\

~\cite{ciacco2026CAF} &
\xmark & \xmark & \cmark &
\cmark &
\cmark &
15 nodes \\
This work &
\xmark & \cmark & \cmark &
\cmark &
\cmark &
30 nodes \\

\hline
\end{tabular}
\caption{
{ Comparison between \textbf{VQA}-based methods and \textbf{QA}-based approaches for the TSP.}
QA approaches are further classified into \textbf{Direct QPU} implementations and \textbf{Hybrid} quantum--classical schemes. 
\textbf{SEC} (subtour elimination constraints) indicates whether subtours are effectively prevented, either explicitly or through implicit mechanisms. 
\textbf{Benchmark instances} specifies whether standard datasets (e.g., TSPLIB) are used for evaluation. 
\textbf{Max size} reports the largest problem instance solved in terms of number of nodes.
}
\label{TabellaQuantum}
\end{table}
}

\section{Problem definition}\label{sec:problem_statement}

The TSP aims to determine a minimum-cost tour that visits each city exactly once and returns to the origin. Let $G = (V, A)$ be a complete graph, where $V = \{1, \dots, n\}$ denotes the set of vertices and $A$ the set of arcs. 
Let $S \subset V$ denote any proper subset of vertices. Each arc $(i,j) \in A$ is associated with a non-negative cost $c_{ij}$ representing the travel cost between vertices $i$ and $j$. We assume a symmetric cost matrix, i.e., $c_{ij} = c_{ji}$ for all $i,j \in V$, and we exclude self-loops, so that arcs of the form $(i,i), \forall i\in V$ are not considered. 
The objective is to find a Hamiltonian cycle of minimum total cost in $G$.

\subsection{Mathematical Formulation}\label{sec:mathematical_formulation}
We introduce the following binary decision variable:
\begin{description}
\centering
    \item[$x_{ij}$] $\begin{cases} 
1 & \text{if arc } (i,j) \text{ is selected in the tour} \\
0 & \text{otherwise}
    \end{cases}$
\end{description}

The TSP can be expressed as the following ILP formulation:

\begin{align}
    &\text{min} \quad \sum_{(i,j)\in A} c_{ij} x_{ij} \label{eq:objective} \\
    &\text{s.t.} \nonumber \\
    &\sum_{i \in V : (i,j)\in A} x_{ij} = 1, \quad \forall j \in V \label{eq:visit_once1} \\
    &\sum_{i \in V : (j,i)\in A} x_{ji} = 1, \quad \forall j \in V \label{eq:visit_once2} \\
    &\sum_{\substack{i \in S}} \sum_{\substack{j \in  S}} x_{ij} \leq |S| - 1 , \quad  S\subset V, |S|\geq 2\label{eq:subtour} \\  
    &x_{ij} \in \{0,1\}, \quad \forall (i,j) \in A \label{eq:binary}
\end{align}

The objective function~\eqref{eq:objective} minimizes the total travel cost of the tour.
Constraints~\eqref{eq:visit_once1} and~\eqref{eq:visit_once2} impose the degree conditions at each vertex: for every city, exactly one incoming arc and one outgoing arc must be selected. As a result, each city is visited exactly once, and the solution consists of a collection of cycles covering all vertices.
Constraints~\eqref{eq:subtour} enforce connectivity by eliminating subtours. 
Finally, constraints~\eqref{eq:binary} define the binary nature of the decision variables.

\subsection{Complexity of the subtour elimination constraints}

Constraints~\eqref{eq:subtour} are subtour elimination constraints whose purpose is to prevent the formation of disconnected cycles in the solution. For any subset of vertices $S \subset V$ with $|S| \ge 2$, the constraints ensures that the vertices in $S$ cannot form an isolated cycle.

Let $n = |V|$. In principle, this constraint must be enforced for every subset $S \subset V$ such that $2 \le |S| \le n-1$.
The total number of such subsets is therefore

\begin{equation}
\sum_{k=2}^{n-1} \binom{n}{k}.
\end{equation}

Using the binomial identity
$
\sum_{k=0}^{n} \binom{n}{k} = 2^n,
$
it follows that

\begin{equation}
\sum_{k=2}^{n-1} \binom{n}{k}
=
2^n - \binom{n}{0} - \binom{n}{1} - \binom{n}{n}
=
2^n - 2 - n.
\end{equation}

Therefore, the number of subtour elimination constraints grows exponentially with the number of vertices and is of order $\mathcal{O}(2^n)$. Consequently, explicitly including all constraints~\eqref{eq:subtour} in the model would lead to a formulation of exponential size, which becomes computationally intractable even for moderately sized instances.

\section{Solution Approach} \label{sec:solution_approach}

This study investigates several strategies for solving the TSP, progressively introducing different modeling choices and algorithmic techniques. In particular, we compare classical exact optimization methods with quantum-inspired and hybrid quantum optimization approaches. The objective is to analyze how modeling decisions and preprocessing strategies influence the tractability of the problem and the computational effort required to obtain feasible tours.

The proposed framework compares classical exact optimization approaches with quantum optimization strategies. In particular, we analyze how the structure of the mathematical model and { the number of nodes affect the performance of the solution process.} 

The overall methodology is organized along two main dimensions:

\begin{itemize}
    \item \textbf{Optimization paradigm}, distinguishing between classical mathematical programming approaches and quantum-inspired formulations;
    \item \textbf{Model reduction strategies}, including both iterative subtour elimination approach and a preprocessing procedure designed to reduce the size of the arc set.
\end{itemize}

The main components of the study are summarized as follows:

\begin{enumerate}
    \item \textbf{Complete ILP formulation {(CILP)}:} as a baseline approach, the TSP is solved using the {classical ILP} 
    formulation introduced in Section~\ref{sec:mathematical_formulation}, which explicitly includes the full set of subtour elimination constraints.

    \item \textbf{{ Cutting-Plane Approach (CPA):} } to address the exponential number of subtour constraints, an iterative constraint generation procedure is adopted. The model is initially solved without subtour elimination constraints, which are then added dynamically whenever violated subtours are detected in intermediate solutions. The details of this procedure are provided in Section~\ref{sec:iterative}.

    \item \textbf{CAF preprocessing:} to reduce the size of the optimization model, a preprocessing step restricts the set of candidate arcs by retaining only the lowest-cost neighbors for each vertex. In particular, for every node the outgoing arcs are sorted according to their travel cost and only a subset of the nearest neighbors is preserved. This filtering procedure reduces the number of decision variables and the density of the optimization model while maintaining a sufficiently connected graph structure that preserves the existence of Hamiltonian cycles. The CAF strategy, originally introduced by \cite{ciacco2026CAF}, is adopted in this work as a preprocessing step to reduce the size of the optimization model.

     \item \textbf{Hybrid quantum optimization:} a hybrid quantum optimization strategy based on the CQM framework is investigated. In this approach, the TSP constraints are modeled explicitly within the CQM formulation, the resulting model is solved using a hybrid solver that integrates classical optimization routines with QA resources.
    
    \item \textbf{Direct quantum-inspired formulation:} the TSP is reformulated as a QUBO problem, where the tour constraints are incorporated into the objective function through penalty terms. The resulting QUBO formulation is embedded and executed directly on a D-Wave QPU.

\end{enumerate}



\subsection{CPA}\label{sec:iterative}

The TSP formulation presented in Section~\ref{sec:mathematical_formulation} contains an exponential number of subtour elimination constraints~\eqref{eq:subtour}. 
Explicitly including all these constraints in the model would lead to a formulation of size $\mathcal{O}(2^n)$, which rapidly becomes computationally intractable even for moderately sized instances.

To address this issue, we adopt a dynamic constraint generation strategy, commonly used in modern branch-and-cut algorithms for the TSP. 
The key idea of this approach is to avoid including the full family of subtour elimination constraints in the initial model. Instead, the optimization problem is solved iteratively by progressively adding only those constraints that are violated by intermediate solutions.

The solution process begins with a restricted formulation of the TSP that includes only the objective function~\eqref{eq:objective}, the degree constraints~\eqref{eq:visit_once1} --~\eqref{eq:visit_once2}, and the binary domain constraints~\eqref{eq:binary}. 
At this stage, the model enforces that each vertex has degree two, but it does not explicitly prevent the formation of disconnected cycles. Consequently, the solution of this relaxed model may consist of multiple subtours covering disjoint subsets of vertices.

After solving the model, the resulting solution is analyzed in order to identify the presence of subtours. 
If the solution consists of more than one connected component, each component corresponds to a subtour defined on a subset of vertices $S \subset V$.
For every detected subtour $S$, the corresponding subtour elimination constraint~(\ref{eq:subtour}) is added to the model. This constraint requires that at least two arcs connect the subset $S$ with its complement $V \setminus S$, thereby preventing the vertices in $S$ from forming an isolated cycle in subsequent solutions.
Once the violated constraints have been added, the model is re-optimized. 

The new solution is again examined to detect possible subtours, and additional constraints are generated whenever necessary. 
This iterative process continues until no violated subtour elimination constraints are found, meaning that the current solution corresponds to a single Hamiltonian cycle visiting all vertices exactly once. The steps of the procedure are depicted in Algorithm \ref{alg:1}.

\begin{algorithm}[H]
\caption{CPA}
\label{alg:1}
\begin{algorithmic}[1]
\Require Graph $G=(V,A)$, travel costs $c_{ij}$
\Ensure Hamiltonian tour

\State Initialize model with:
\Statex \quad objective function~\eqref{eq:objective}
\Statex \quad degree constraints~\eqref{eq:visit_once1} --~\eqref{eq:visit_once2}
\Statex \quad binary constraints~\eqref{eq:binary}

\Repeat
    \State Solve the current ILP model
    \State Extract the solution graph induced by variables $x_{ij}$
    \State Identify connected components (subtours)

    \If{more than one component exists}
        \For{each subtour $S \subset V$}
            \State Add subtour elimination constraint~\eqref{eq:subtour}
        \EndFor
    \EndIf

\Until{the solution consists of a single Hamiltonian cycle}

\State \Return final tour
\end{algorithmic}
\end{algorithm}

From a computational standpoint, this incremental strategy significantly improves the tractability of the model. 
Rather than handling the full exponential family of subtour elimination constraints simultaneously, the solver dynamically generates only the constraints that are required to eliminate infeasible intermediate solutions encountered during the optimization process. 
This approach preserves the correctness of the formulation while maintaining a manageable model size.

\section{Computational study} \label{sec:experiments} 
This section is structured into two main parts. The first part, \textit{Computational setup and solver configuration}, outlines the hardware and software environment together with the configuration of the solvers employed in the experiments. The second part, \textit{Numerical results}, reports and discusses the computational findings, emphasizing the performance of the proposed models and solution approaches.
\subsection{Computational setup and solver configuration}
All computational experiments are conducted on a Windows 10 machine equipped with an Intel processor (Family 6, Model 126, Stepping 5), featuring four physical cores (eight logical threads) and 15.6 GB of RAM. The implementation is developed in Python~3. For exact optimization experiments, we employ Gurobi Optimizer version~11.0.1 \cite{gurobi-doc}.

Furthermore, we use the \texttt{dimod.cqm\_to\_bqm} method from the \texttt{Ocean SDK} library to convert the ILP model, which includes constraints, into a Binary Quadratic Model (BQM). This conversion transforms the constraints into penalty terms, resulting in an unconstrained model that can be further processed as a QUBO formulation for quantum optimization. It is important to note that this conversion alters the structure of the problem. Due to the introduction of bias and penalty terms, the solutions obtained from the ILP and QUBO formulations may exhibit different energy values.


We employ two distinct approaches for QA using resources provided by D-Wave. 

The first approach utilizes a D-Wave quantum annealer, specifically the \texttt{Advantage2\_system1.13} solver, accessed through the \texttt{DWaveSampler} combined with the \texttt{EmbeddingComposite} to automatically handle minor-embedding.

In this approach, the total computational time can be decomposed into two main components:
\begin{itemize}
    \item \textbf{Build Time}: includes all operations required to construct the optimization problem;
    
    \item \textbf{Computation Time}: corresponds to the interaction with the QPU and includes all operations required to obtain samples from the device. It follows a structured pipeline and can be decomposed as follows:
    
    \begin{itemize}
        \item \textbf{Conversion time}: time required to transform the ILP formulation into a QUBO model suitable for quantum processing;
        
        \item \textbf{QPU time}: time required to interact with the quantum hardware, including:
        \begin{itemize}
            \item \textbf{Overhead time}: time required to submit the problem to the QPU, including communication latency and queue waiting time;
            
            \item \textbf{Solver time}: time spent by the QPU to generate samples, which includes:
            \begin{itemize}
                \item \textbf{Annealing time}: duration of each annealing process;
                \item \textbf{Readout time}: time required to measure the qubits.
            \end{itemize}
        \end{itemize}
        
        \item \textbf{Decoding time}: time required to map the QUBO solution back to the original ILP formulation.
    \end{itemize}
\end{itemize}

It is important to note that the solver time is inherently based on repeated sampling: each sample corresponds to an independent annealing run. Consequently, the total solver time depends both on the number of samples (reads) and on the duration of each individual annealing process.
More precisely, the solver time can be approximated as:
\begin{equation}
T_{\text{solver}} \approx  \texttt{num\_reads} \cdot (\texttt{annealing\_time} + \texttt{readout\_time}),
\end{equation}
where:
\begin{itemize}
    \item \texttt{num\_reads} is the number of independent annealing runs (sampling depth);
    \item \texttt{annealing\_time} is the duration of a single annealing process;
\item \texttt{readout\_time} denotes the time required for qubit measurement, i.e., the process of retrieving the computed solutions from the quantum hardware.
\end{itemize}
The readout time is fixed by the hardware, whereas the annealing time can be controlled by the user. Together with the number of reads, these parameters define a trade-off between solution quality and solver time.
Due to hardware constraints, the total solver time must satisfy:
\begin{equation}
\texttt{num\_reads} \cdot (\texttt{annealing\_time} + \texttt{readout\_time}) \leq T_{\max}, 
\label{numreads}
\end{equation}
where $T_{\max}$ represents the maximum allowed solver time per submission ($10^6\,\mu s$).

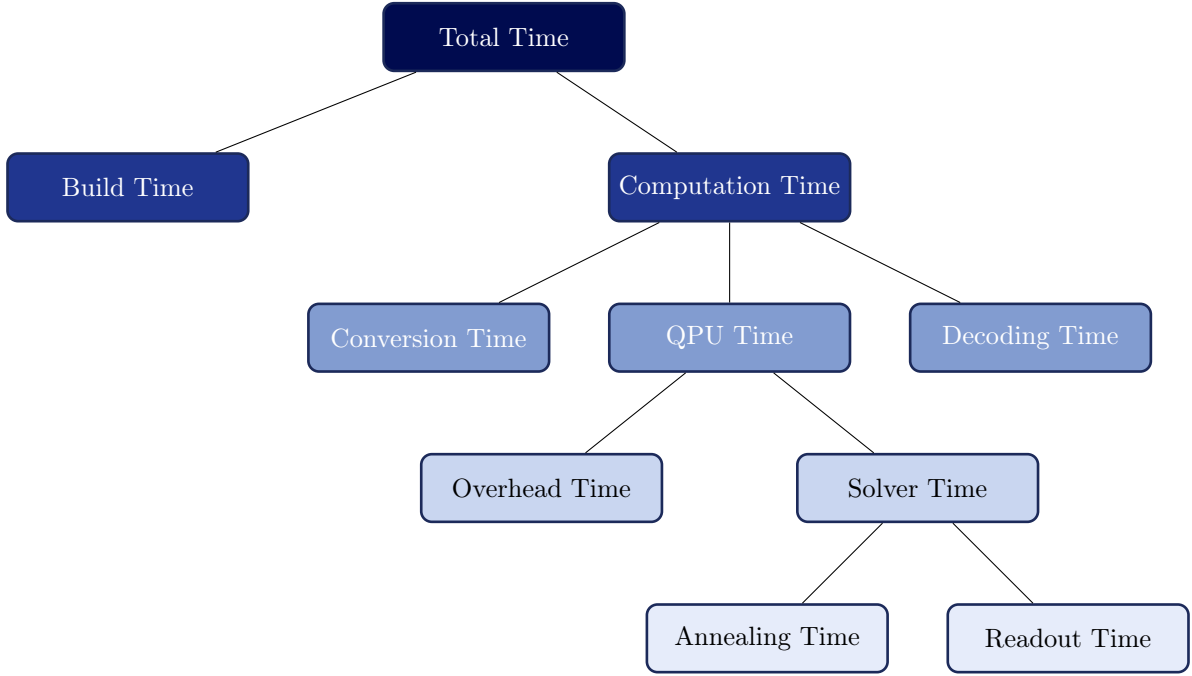
\begin{figure}[h]
\centering
\begin{tikzpicture}[
node distance=1.5cm and 2.5cm,
    every node/.style={
        draw,
        rectangle,
        rounded corners,
        align=center,
        minimum width=3.2cm,
        minimum height=0.9cm,
        font=\small,
        line width=1pt
    }
]
\node[fill=myblue, draw=myborder, text=white] (root) at (0,0) {Total Time};

\node[fill=myblue2, draw=myborder, text=white] (build) at (-5,-2) {Build Time};
\node[fill=myblue2, draw=myborder, text=white] (comp) at (3,-2) {Computation Time};

\node[fill=myblue3, draw=myborder, text=white] (conv) at (-1,-4) {Conversion Time};
\node[fill=myblue3, draw=myborder, text=white] (qpu) at (3,-4) {QPU Time};
\node[fill=myblue3, draw=myborder, text=white] (dec) at (7,-4) {Decoding Time};

\node[fill=myblue4, draw=myborder] (over) at (0.5,-6) {Overhead Time};
\node[fill=myblue4, draw=myborder] (solv) at (5.5,-6) {Solver Time};

\node[fill=myblue5, draw=myborder] (ann) at (3.5,-8) {Annealing Time};
\node[fill=myblue5, draw=myborder] (read) at (7.5,-8) {Readout Time};

\draw (root) -- (build);
\draw (root) -- (comp);

\draw (comp) -- (conv);
\draw (comp) -- (qpu);
\draw (comp) -- (dec);

\draw (qpu) -- (over);
\draw (qpu) -- (solv);

\draw (solv) -- (ann);
\draw (solv) -- (read);

\end{tikzpicture}
\caption{
Decomposition of the overall computational time into its main components, highlighting their hierarchical organization and the contribution of each phase in the direct QA workflow.
}
\label{fig:time_decomposition}
\end{figure}

This observation highlights a fundamental trade-off in direct QA: increasing the number of reads improves the exploration of the solution space, whereas increasing the annealing time enhances the quality of each individual sample. However, both parameters directly impact the total solver time and must therefore be carefully balanced.
Figure~\ref{fig:time_decomposition} provides a visual representation of the different components contributing to the overall computational time, highlighting their hierarchical organization.

In our experiments, the annealing time is fixed to $100\,\mu s$, while the number of reads is determined according to the following rule:
\begin{equation} 
\texttt{num\_reads} = \min\left(n_{\text{reads}},\; \texttt{num\_reads}_{\max} \right).
\end{equation}

The upper bound $\texttt{num\_reads}_{\max}$ is derived from inequality~(\ref{numreads}), ensuring compliance with the QPU time limitations. In particular, the readout time is approximately $115\,\mu s$, as determined by the hardware and not user-controlled, while the annealing time is fixed to $100\,\mu s$. Therefore, the time required per sample is approximately $215\,\mu s$, leading to
\begin{equation}
\texttt{num\_reads}_{\max} \approx \frac{T_{max}}{215\,\mu s},
\end{equation}
The term $n_{\text{reads}}$ represents a target number of reads, defined differently depending on the adopted strategy:

\begin{itemize}
    \item \textbf{CPA:}
    \begin{equation}
    n_{\text{reads}} = n_{\text{start}} + 100 \cdot |\mathcal{C}|,
    \end{equation}
    where $n_{\text{start}} = 1000$ and $|\mathcal{C}|$ denotes the total number of subtour elimination constraints added across all previous iterations.

    \item \textbf{CILP:}
    \begin{equation}
    n_{\text{reads}} = n_{\text{start}} + 100 \cdot |\mathcal{C}|_{\max},
    \end{equation}
where $n_{\text{start}} = 1000$ and $|\mathcal{C}|_{\max}$ denotes the maximum number of subtour elimination constraints observed during the CPA, which scales with the number of customers in the problem.
\end{itemize}

\begin{figure}[!ht]
    \centering
    \includegraphics[width=0.55\linewidth]{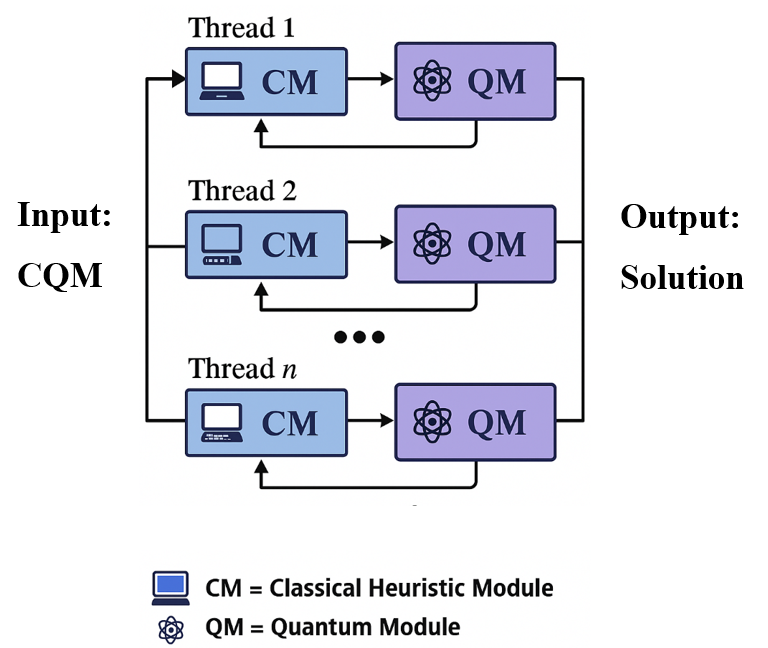}
    \caption{General structure of D-Wave’s hybrid solvers, from \cite{ciacco2025steinerv2}.}
    \label{fig:hybrid_architecture}
\end{figure}

The CQM formulation is implemented using D-Wave's Ocean SDK, specifically \texttt{dimod} version~0.12.18 and \texttt{dwave-system} version~1.28.0. All problem instances are solved through D-Wave's hybrid solver service for CQMs, identified as \texttt{hybrid\_constrained\_quadratic\_model\_version1p}. 
We utilize the \texttt{LeapCQMHybrid} solver, a proprietary hybrid optimization tool developed by D-Wave Systems within the Leap cloud platform. This framework integrates classical and quantum computing resources into a unified optimization pipeline. As described in recent studies on hybrid quantum-classical architectures \cite{osaba2025d, bertuzzi2024evaluation, willsch2022benchmarking}, the Hybrid Solver Service operates through a modular asynchronous workflow that combines classical heuristics with QA.
Each problem instance is decomposed into parallel branches, where a Classical Module (CM) performs large-scale exploration using metaheuristics such as local search, tabu search, and simulated annealing, while a Quantum Module (QM) performs quantum-guided refinement through annealing cycles executed on the \texttt{D-Wave Advantage2\_system1.4} QPU. The two modules interact asynchronously: the CM generates candidate solutions and invokes the QM when diversification is required, while quantum samples are returned to guide further classical search.
Figure~\ref{fig:hybrid_architecture} illustrates this hybrid interaction between classical and quantum components.

Due to the proprietary nature of the \texttt{LeapCQMHybrid} solver, detailed information about its internal quantum routines, resource allocation, and the exact number of qubits used for each instance is not publicly disclosed. Additional technical details and benchmark analyses can be found in D-Wave’s official documentation \cite{dwave_cqm_solver, dwave_cqm_parameters}. 

\subsection{Numerical results}  \label{sec:results} 
We present the computational experiments carried out to evaluate the performance of the proposed solution approaches. 

All experiments are conducted on benchmark instances from the \texttt{berlin52.tsp} instances from the TSPLIB benchmark library \cite{reinelt1991tsplib}, a widely adopted dataset in the optimization literature for evaluating algorithms for the TSP. The use of these standardized instances allows for reproducible experiments and meaningful comparisons with previously published results.

The computational study is organized into three main parts. First, we analyze the impact of the proposed reduction strategies on the complexity of the optimization model. Second, we evaluate the performance of classical optimization approaches. Finally, we investigate quantum optimization methods, including both direct QPU-based formulations and hybrid quantum--classical approaches.

\subsubsection{Impact of the proposed solution approach on model complexity}
Table~\ref{tab:complessità} reports the impact of the CAF preprocessing and the CPA on the size of the optimization model, in terms of number of variables and constraints. 

For each instance size $V$, the table is structured into two main sections: the CILP and the CPA. Within each part, results are presented both without CAF and with CAF, referring respectively to the absence and the application of the preprocessing strategy. For each configuration, the table reports the number of decision variables (Var) and the number of constraints (Constr).
Additionally, the table includes the percentage reduction in the number of variables achieved through CAF, computed as $(\text{Var}_{\text{without CAF}} - \text{Var}_{\text{with CAF}})/\text{Var}_{\text{without CAF}}\times 100$, and the percentage reduction in the number of constraints obtained by comparing CPA with CILP, computed as $(\text{Constr}_{\text{CILP}} - \text{Constr}_{\text{CPA}})/\text{Constr}_{\text{CILP}}\times 100$.

\begin{table}[h]
\centering
\resizebox{0.8\columnwidth}{!}{
\begin{tabular}{c|cc|cc|cc|cc|>{\centering\arraybackslash}p{2.5cm}|>{\centering\arraybackslash}p{2.5cm}}
\toprule
\multirow{3}{*}{\textbf{V}} 
& \multicolumn{4}{c|}{\textbf{CILP}} 
& \multicolumn{4}{c|}{\textbf{CPA}} 
& \multirow{3}{*}{\textbf{Var Red.}} 
& \multirow{3}{*}{\textbf{Constr. Red.}} \\
\cmidrule(lr){2-5} \cmidrule(lr){6-9}
& \multicolumn{2}{c|}{\textbf{Without CAF}} 
& \multicolumn{2}{c|}{\textbf{With CAF}} 
& \multicolumn{2}{c|}{\textbf{Without CAF}} 
& \multicolumn{2}{c|}{\textbf{With CAF}} & \multicolumn{1}{c|}{} &  \\
\cmidrule(lr){2-3} \cmidrule(lr){4-5} \cmidrule(lr){6-7} \cmidrule(lr){8-9}
& Var & Constr 
& Var & Constr 
& Var & Constr 
& Var & Constr 
& \multicolumn{1}{c|}{} &  \\
\midrule
5  & 20 & 35    & 18 & 35    & 20 & 14 & 18 & 14 & 10\% & 60\% \\
6  & 30 & 68    & 24 & 68    & 30 & 17 & 24 & 17 & 20\% & 75\% \\
7  & 42 & 133   & 34 & 133   & 42 & 20 & 34 & 20 & 19\% & 85\% \\
8  & 56 & 262   & 42 & 262   & 56 & 21 & 42 & 21 & 25\% & 92\% \\
9  & 72 & 519   & 58 & 519   & 72 & 22 & 58 & 24 & 19\% & 95\% \\
10 & 90 & 1032  & 64 & 1032  & 90 & 27 & 64 & 27 & 29\% & 97\% \\
11 & 110 & 2057 & 88 & 2057  & 110 & 31 & 88 & 31 & 20\% & 98\% \\
12 & 132 & 4106 & 96 & 4106  & 132 & 36 & 96 & 36 & 27\% & 99\% \\
13 & 156 & 8203 & 118 & 8203 & 156 & 39 & 118 & 39 & 24\% & 100\% \\
14 & 182 & 16396 & 126 & 16396 & 182 & 44 & 126 & 46 & 31\% & 100\% \\
15 & 210 & 32781 & 154 & 32781 & 210 & 46 & 154 & 46 & 27\% & 100\% \\
20 & 380 & 1048594 & 262 & 1048594 & 380 & 61 & 262 & 61 & 31\% & 100\% \\
25 & 600 & -- & 420 & -- & 600 & 83 & 420 & 83 & 30\% & -- \\
30 & 870 & -- & 562 & -- & 870 & 99 & 562 & 99 & 35\% & -- \\
35 & 1190 & -- & 806 & -- & 1190 & 91 & 806 & 91 & 32\% & -- \\
40 & 1560 & -- & 1056 & -- & 1560 & 104 & 1056 & 104 & 32\% & -- \\
45 & 1980 & -- & 1358 & -- & 1980 & 116 & 1358 & 116 & 31\% & -- \\
\bottomrule
\end{tabular}
}
\caption{{ Model complexity comparison between CILP and CPA, with and without CAF preprocessing.} The entries ``--'' indicate instances for which the CILP cannot be constructed within the imposed time limit of 1000 seconds allocated exclusively to the model generation phase, due to its prohibitive size. }
\label{tab:complessità}
\end{table}
\color{black}

The results show that the CAF preprocessing consistently reduces the number of decision variables across all tested instances. On average, the reduction is approximately 26\%, increasing to about 32\% for $n \geq 20$. This trend indicates that CAF becomes more effective as the problem size grows, making it particularly suitable for larger instances. Overall, CAF efficiently filters out non-promising arcs while preserving the structure of the problem. Moreover, the reduction remains stable across different instance sizes, highlighting the robustness of the preprocessing strategy.

A more pronounced effect is observed in the number of constraints. The CPA drastically reduces the number of constraints compared to the CILP. While the CILP requires an exponential number of subtour elimination constraints, the CPA introduces only those that are violated during the solution process.
As a result, the reduction in the number of constraints rapidly approaches 100\% as the problem size increases. Even for medium-sized instances, the number of constraints is reduced by more than 95\%, while for larger instances the reduction becomes nearly total. This demonstrates the crucial role of the CPA in mitigating the combinatorial explosion associated with subtour elimination constraints.

Thus, the results highlight the complementary nature of the two techniques: CAF reduces the number of variables, leading to a more compact search space, while the CPA significantly limits the number of constraints, improving the tractability of the model. Their combined use is therefore essential for addressing larger instances, particularly in the context of quantum and hybrid optimization approaches.

\subsubsection{Classical Optimization Results}\label{sec:classic}

Table~\ref{tab:milp_results} reports the computational results obtained for the CILP and the CPA, both with and without CAF preprocessing.

The table presents results for TSP instances with an increasing number of vertices $V$, allowing a comparison between the performance of the CILP formulation and the CPA approach. The results are provided both in the absence and in the presence of the CAF preprocessing procedure.
Each row corresponds to a problem instance characterized by the number of vertices $V$. For each configuration, the table reports the objective function value (OF) returned by the solver, the total computational time in seconds (Time), which includes both model construction and solution phases, the solution time in seconds (Time Solve) required by the solver to optimize the model, and the optimality gap (GAP), expressed as a percentage with respect to the best solution found by the solver.

\color{black}

\begin{table}[h]
\centering
\resizebox{\columnwidth}{!}{
\begin{tabular}{c|cccc|cccc|cccc|cccc}
\toprule
\multirow{3}{*}{\textbf{V}} 
& \multicolumn{8}{c|}{\textbf{CILP}} 
& \multicolumn{8}{c}{\textbf{CPA}} \\
\cmidrule(lr){2-9} \cmidrule(lr){10-17}
& \multicolumn{4}{c|}{\textbf{Without CAF}} 
& \multicolumn{4}{c|}{\textbf{With CAF}} 
& \multicolumn{4}{c|}{\textbf{Without CAF}} 
& \multicolumn{4}{c}{\textbf{With CAF}} \\
\cmidrule(lr){2-5} \cmidrule(lr){6-9} \cmidrule(lr){10-13} \cmidrule(lr){14-17}
 & OF & Time & Time Solve & GAP 
 & OF & Time & Time Solve & GAP
 & OF & Time & Time Solve & GAP
 & OF & Time & Time Solve & GAP \\
\midrule

    5     & 2314,55 & 0,00 & 0,00 & 0\%   & 2314,55 & 0,08 & 0,04 & 0\%   & 2314,55 & 0,01  & 0,00  & 0\%   & 2314,55 & 0,01  & 0,00  & 0\% \\
    6     & 2315,15 & 0,00 & 0,00 & 0\%   & 2323,20 & 0,00 & 0,00 & 0\%   & 2315,15 & 0,01  & 0,01  & 0\%   & 2323,20 & 0,01  & 0,00  & 0\% \\
    7     & 2321,39 & 0,01 & 0,00 & 0\%   & 2321,39 & 0,01 & 0,00 & 0\%   & 2321,39 & 0,01  & 0,01  & 0\%   & 2321,39 & 0,01  & 0,01  & 0\% \\
    8     & 2550,94 & 0,01 & 0,00 & 0\%   & 2550,94 & 0,02 & 0,01 & 0\%   & 2550,94 & 0,01  & 0,01  & 0\%   & 2550,94 & 0,01  & 0,01  & 0\% \\
    9     & 2820,38 & 0,04 & 0,02 & 0\%   & 2874,44 & 0,02 & 0,01 & 0\%   & 2820,38 & 0,01  & 0,00  & 0\%   & 2874,44 & 0,01  & 0,01  & 0\% \\
    10    & 2826,50 & 0,09 & 0,03 & 0\%   & 2826,50 & 0,08 & 0,04 & 0\%   & 2826,50 & 0,01  & 0,01  & 0\%   & 2826,50 & 0,01  & 0,01  & 0\% \\
    11    & 4038,44 & 0,21 & 0,09 & 0\%   & 4038,44 & 0,16 & 0,08 & 0\%   & 4038,44 & 0,01  & 0,01  & 0\%   & 4038,44 & 0,01  & 0,01  & 0\% \\
    12    & 4056,68 & 0,50 & 0,17 & 0\%   & 4056,68 & 0,30 & 0,17 & 0\%   & 4056,68 & 0,02  & 0,01  & 0\%   & 4056,68 & 0,02  & 0,01  & 0\% \\
    13    & 4564,46 & 1,25 & 0,47 & 0\%   & 4564,46 & 0,81 & 0,49 & 0\%   & 4564,46 & 0,02  & 0,01  & 0\%   & 4564,46 & 0,02  & 0,01  & 0\% \\
    14    & 4946,85 & 3,26 & 1,32 & 0\%   & 4965,33 & 1,65 & 1,15 & 0\%   & 4946,85 & 0,03  & 0,02  & 0\%   & 4965,33 & 0,03  & 0,02  & 0\% \\
    15    & 4967,30 & 8,20 & 2,97 & 0\%   & 4967,30 & 4,70 & 3,28 & 0\%   & 4967,30 & 0,02  & 0,02  & 0\%   & 4967,30 & 0,02  & 0,02  & 0\% \\
    20    & 5270,86 & 670,99 & 264,14 & 0\%   & 5270,86 & 316,13 & 257,66 & 0\%   & 5270,86 & 0,03  & 0,02  & 0\%   & 5270,86 & 0,03  & 0,02  & 0\% \\
    25    & -     & -     & -     & -     & -     & -     & -     & -     & 5460,94 & 0,14  & 0,08  & 0\%   & 5460,94 & 0,09  & 0,05  & 0\% \\
    30    & -     & -     & -     & -     & -     & -     & -     & -     & 6146,65 & 0,34  & 0,27  & 0\%   & 6146,65 & 0,27  & 0,21  & 0\% \\
    35    & -     & -     & -     & -     & -     & -     & -     & -     & 6557,12 & 0,13  & 0,10  & 0\%   & 6557,12 & 0,10  & 0,08  & 0\% \\
    40    & -     & -     & -     & -     & -     & -     & -     & -     & 6652,63 & 0,15  & 0,12  & 0\%   & 6652,63 & 0,12  & 0,10  & 0\% \\
    45    & -     & -     & -     & -     & -     & -     & -     & -     & 6887,37 & 0,14  & 0,07  & 0\%   & 6887,37 & 0,12  & 0,08  & 0\% \\

\bottomrule
\end{tabular}
}
\caption{Computational results for TSP instances with increasing number of vertices $V$, comparing { the CILP and CPA formulations, with and without CAF preprocessing.}
The entries ``--'' indicate instances for which the {CILP model cannot be constructed within the imposed time limit of 1000 seconds.}}
\label{tab:milp_results}
\end{table}

The CILP shows a rapid growth in computational time as the problem size increases. While small instances ($V \leq 12$) are solved within fractions of a second, the total runtime increases substantially for larger instances. In particular, for $V=15$ the solution time already reaches several seconds, and for $V=20$ it becomes prohibitive. Beyond this size, the CILP cannot be constructed within the imposed time limits, highlighting the intrinsic scalability limitations of the full model.
The introduction of CAF preprocessing consistently improves the performance of the CILP. For medium-sized instances, CAF significantly reduces the total runtime, roughly halving the computational effort (e.g., from 670,99 to 316,13 seconds for $V=20$). This confirms that reducing the number of variables leads to a more tractable optimization problem, although it is not sufficient to overcome the exponential growth of constraints.

A markedly different behavior is observed for the CPA. This formulation exhibits extremely low computational times across all tested instances, remaining below a fraction of a second even for the largest instances ($V=45$). The results clearly demonstrate that dynamically generating subtour constraints avoids the combinatorial explosion associated with the CILP.
Moreover, the CPA scales effectively to larger instances where the CILP is no longer applicable. Even for $V \geq 25$, where the CILP cannot be constructed, the CPA continues to produce optimal solutions with negligible computational effort.
The impact of CAF on the CPA is still beneficial. Since the number of constraints is already drastically reduced by the CPA, CAF still provides a consistent reduction in runtime. For larger instances, this effect becomes more evident: for example, at $V=30$ the total time decreases from 0,34 to 0,27 seconds, and at $V=45$ from 0,14 to 0,12 seconds. Therefore, for larger instances, CAF provides a noticeable reduction in computational time also in the CPA, with improvements ranging from approximately 14\% to 36\%, and an average reduction of about 23\%. While its impact is negligible for small instances due to already minimal runtimes, CAF becomes increasingly beneficial as the problem size grows, further enhancing the efficiency of the CPA.

Overall, these results highlight that the combination of the CPA with CAF preprocessing yields the best overall performance, enabling the solution of significantly larger instances with minimal computational cost. In particular, the CPA reduces the number of constraints by more than 95\% already for medium-sized instances and up to nearly 100\% for larger ones, effectively eliminating the exponential growth associated with the CILP. 
At the same time, CAF preprocessing reduces the number of decision variables by approximately 20\%--30\%, with even higher reductions (up to about 32\%) for larger instances. The combined effect of these two strategies leads to a drastic simplification of the optimization model, allowing instances with up to $V=45$ to be solved in a fraction of a second, whereas the CILP becomes intractable already at $V=20$.

\subsubsection{Quantum Optimization Results}
We analyze the performance of quantum optimization approaches. The analysis is structured into two parts. First, we consider direct quantum formulations, executed directly on the QPU, where the problem is encoded into a QUBO model and solved with QA. Second, we investigate hybrid quantum--classical approaches, where classical optimization procedures are combined with QA techniques to enhance solution quality and scalability.
All experiments are performed using the CAF preprocessing strategy. This choice is supported both by the results in \cite{ciacco2026CAF} and by our findings in Section~\ref{sec:classic}, which consistently demonstrate that CAF effectively reduces the model size while preserving solution quality. Furthermore, each test instance is executed over five independent runs in order to account for variability in the solution process and to obtain more robust performance evaluations.
For the CPA, a maximum number of iterations is imposed for each run due to limited computational resources. This limit is defined as a function of the problem size $n$, and is set to $\lceil n/2 \rceil$. This choice is motivated by the structure of the CPA. In the worst-case scenario, the solution of the relaxed model (i.e., without subtour elimination constraints) may consist of multiple disjoint cycles. Since each subtour must contain at least two vertices, the maximum possible number of disjoint subtours is bounded by $n/2$.

\subsubsubsection{Direct QPU-Based results}\label{sec:direct}
The results obtained using the direct QPU-based approach are reported in Table~\ref{tab:direct}. 

The table presents results for TSP instances with an increasing number of vertices $V$, comparing the performance of the CILP and CPA approaches under a direct QPU-based solving framework. 
The table reports the following metrics: the average objective function value returned by the solver (OF$_{\mathrm{avg}}$) and its standard deviation (OF$_{\mathrm{dev}}$); the average total computational time in seconds (Time$_{\mathrm{avg}}$) and its standard deviation (Time$_{\mathrm{dev}}$), where for the CPA this value is computed as the sum of the computational times over all iterations within each run; the average model construction time (Build$_{\mathrm{avg}}$) and its standard deviation (Build$_{\mathrm{dev}}$); the average total computation time (Comp$_{\mathrm{avg}}$) and its standard deviation (Comp$_{\mathrm{dev}}$); the average solver time (Solve$_{\mathrm{avg}}$) and its standard deviation (Solve$_{\mathrm{dev}}$), where for the CPA this value is computed as the sum of the solver times over all iterations within each run; the average number of generated cuts (Cuts$_{\mathrm{avg}}$) and its standard deviation (Cuts$_{\mathrm{dev}}$); the average number of iterations of the subtour elimination procedure in the CPA (Iters$_{\mathrm{avg}}$) and its standard deviation (Iters$_{\mathrm{dev}}$); the optimality gap (GAP), expressed as a percentage with respect to the optimal solution reported in Table~\ref{tab:milp_results}; and the percentage of feasible solutions over the total number of runs (Feas).
\color{black}

For the CILP, experiments could only be performed for instances with $V \leq 8$, as QPU access limitations prevented the execution of larger instances ($V \geq 9$). Consequently, no results are available beyond this size.
For the CPA, experiments were carried out up to $V = 12$. However, no feasible solutions were found for $V = 11$ and $V = 12$, and the analysis was therefore terminated at this point.

\begin{table}[h]
\centering

\resizebox{\columnwidth}{!}{
\begin{tabular}{
>{\centering\arraybackslash}p{0.8cm}|
*{11}{>{\centering\arraybackslash}p{1.8cm}}|
>{\centering\arraybackslash}p{1.5cm}|
>{\centering\arraybackslash}p{1.5cm}
}
\toprule
\multirow{2}{*}{\textbf{V}} &
\multicolumn{13}{c}{\textbf{CILP (With CAF)}} \\
\cmidrule(lr){2-14}
& OF$_{avg}$ & OF$_{dev}$
& Time$_{avg}$ & Time$_{dev}$
& Build$_{avg}$ & Build$_{dev}$
& Comp$_{avg}$ & Comp$_{dev}$
& Solve$_{avg}$ & Solve$_{dev}$
& Cuts$_{avg}$
& \textbf{GAP (\%)}
& \textbf{Feas (\%)} \\
\midrule
    5     & 2613,04 & 389,97 & 2,52  & 0,12  & 0,00  & 0,00  & 2,52  & 0,12  & 0,34  & 0,01  & 10    & 13\%  & 80\% \\
    6     & 1836,79 & 920,92 & 6,12  & 2,38  & 0,01  & 0,00  & 6,11  & 2,38  & 0,37  & 0,01  & 23    & -     & 0\% \\
    7     & 2326,21 & 645,38 & 54,48 & 18,80 & 0,12  & 0,24  & 54,37 & 18,65 & 0,45  & 0,02  & 53    & -     & 0\% \\
    8     & 2605,42 & 674,92 & 236,15 & 86,88 & 0,03  & 0,02  & 236,12 & 86,87 & 0,47  & 0,01  & 113   & -     & 0\% \\

\bottomrule
\end{tabular}
}

\vspace{0.5cm}

\resizebox{\columnwidth}{!}{
\begin{tabular}{
>{\centering\arraybackslash}p{0.8cm}|
*{14}{>{\centering\arraybackslash}p{1.2cm}}|
>{\centering\arraybackslash}p{1.5cm}| >{\centering\arraybackslash}p{1.5cm}
}
\toprule
\multirow{2}{*}{\textbf{V}} &
\multicolumn{15}{c}{\textbf{CPA (With CAF)}} \\
\cmidrule(lr){2-17}
& OF$_{avg}$ & OF$_{dev}$
& Time$_{avg}$ & Time$_{dev}$
& Build$_{avg}$ & Build$_{dev}$
& Comp$_{avg}$ & Comp$_{dev}$
& Solve$_{avg}$ & Solve$_{dev}$
& Iters$_{avg}$ & Iters$_{dev}$
& Cuts$_{avg}$ & Cuts$_{dev}$
& \textbf{GAP (\%)}
& \textbf{Feas (\%)} \\
\midrule
    5     & 2314,55 & 0,00  & 7,15  & 0,46  & 0,01  & 0,00  & 7,14  & 0,47  & 0,86  & 0,01  & 3     & 0     & 4     & 0     & 0\%   & 100\% \\
    6     & 2323,20 & 0,00  & 8,08  & 1,64  & 0,01  & 0,00  & 8,08  & 1,63  & 0,89  & 0,01  & 3     & 0     & 5     & 0     & 0\%   & 100\% \\
    7     & 2324,91 & 4,85  & 9,46  & 1,27  & 0,01  & 0,00  & 9,45  & 1,26  & 1,02  & 0,17  & 3     & 0     & 6     & 0     & 0\%   & 100\% \\
    8     & 2573,25 & 12,14 & 7,29  & 2,61  & 0,01  & 0,00  & 7,28  & 2,60  & 0,73  & 0,34  & 2     & 1     & 4     & 2     & 1\%   & 100\% \\
    9     & 3485,67 & 401,76 & 19,18 & 12,49 & 0,02  & 0,02  & 19,16 & 12,48 & 1,34  & 0,84  & 4     & 2     & 8     & 6     & 21\%  & 40\% \\
    10    & 4486,02 & 1147,12 & 34,37 & 20,06 & 0,03  & 0,01  & 34,34 & 20,05 & 1,77  & 0,68  & 4     & 1     & 13    & 6     & 59\%  & 40\% \\
    11    & 7338,70 & 1017,32 & 1376,20 & 2785,10 & 0,05  & 0,01  & 1376,14 & 2785,10 & 2,90  & 0,30  & 6     & 0     & 22    & 5     & -  & 0\% \\
    12    & 7894,96 & 987,54 & 198,71 & 62,86 & 0,06  & 0,01  & 198,65 & 62,86 & 2,94  & 0,31  & 6     & 0     & 24    & 4     & -  & 0\% \\
\bottomrule
\end{tabular}
}
\caption{{ Computational results for TSP instances with increasing number of vertices $V$ under a direct QPU-based solving framework, comparing the performance of the CILP and CPA approaches.}
The entries ``--'' indicate instances for which no feasible solutions were obtained, and therefore the optimality gap could not be computed. 
}
\label{tab:direct}
\end{table}
The results reported in Table~\ref{tab:direct} highlight significant differences between the CILP and CPA when solved using a direct QPU-based approach. In particular, for the CILP, solutions are obtained only for the smallest instance size ($V = 5$), with a feasibility rate of $80\%$ and an optimality gap of $13\%$ with respect to the Gurobi optimum. For larger instances ($V \geq 6$), no feasible solutions are found, indicating that the direct encoding of the CILP quickly becomes intractable for the QPU. In contrast, the CPA exhibits significantly better performance. For instances with $V \leq 7$, all runs produce feasible solutions and achieve optimality ($0\%$ gap). For $V = 8$, the method still guarantees $100\%$ feasibility, with a very small optimality gap of only $1\%$, demonstrating strong robustness and solution quality. As the instance size increases further, the performance of the CPA gradually degrades. For $V = 9$, feasible solutions are obtained in $40\%$ of the runs, with an average optimality gap of $21\%$. For $V = 10$, the feasibility rate remains at $40\%$, but the optimality gap increases significantly to $59\%$, reflecting the growing difficulty of the problem.
Finally, for $V = 11$ and $V = 12$, no feasible solutions are obtained. This marks the practical scalability limit of the CPA under the direct QPU-based approach.
This suggests that the CPA is more robust in guiding the search toward feasible and competitive solutions when leveraging QA.

From a computational perspective, the CILP exhibits lower execution times for very small instances (e.g., $V=5$). However, its computational time grows rapidly with the problem size, reaching over 200 seconds already at $V=8$, and becomes infeasible beyond this point. 
In contrast, the CPA shows a more controlled growth in computational time for small and medium instances. Although it requires multiple QPU calls, resulting in higher total times for small instances, it scales more effectively. This behavior is reflected in both the computation time and solver time, which remain manageable up to $V=10$. However, for larger instances ($V \geq 11$), the total time increases significantly due to the accumulation of multiple iterations, without yielding feasible solutions.

The number of generated cuts provides further insight into the behavior of the two formulations. 
For the CILP, all subtour elimination constraints are included a priori in the model. As a consequence, the number of cuts grows rapidly with the instance size, increasing from 10 for $V=5$ to 113 for $V=8$. This results in a very large and dense QUBO formulation, which quickly becomes intractable for the QPU and explains the inability to find feasible solutions for larger instances. In contrast, the CPA introduces cuts dynamically, adding only the subtour elimination constraints that are actually violated by the current solution. As a result, the number of cuts remains significantly lower and grows more gradually. For small instances ($V \leq 7$), the average number of cuts ranges between 4 and 6, while for larger instances it increases to 13 for $V=10$ and up to 24 for $V=12$. 

Overall, the experimental results clearly highlight the limitations of the CILP when implemented in a direct QPU-based setting. While it is able to produce feasible solutions for very small instances, the rapid growth in the number of constraints leads to large and dense QUBO models, which quickly become intractable for the quantum hardware.
In contrast, the CPA demonstrates significantly better scalability and robustness. By dynamically introducing subtour elimination constraints, it maintains a more compact model and is able to consistently produce feasible and high-quality solutions for a wider range of instance sizes. This advantage is particularly evident in terms of feasibility, optimality gap, and controlled growth of computational time.
However, the results also reveal the intrinsic limitations of direct QA. As the problem size increases, both formulations eventually fail to produce feasible solutions, highlighting the impact of hardware constraints and the complexity of the underlying combinatorial problem.

\subsubsubsection{Hybrid quantum--classical results}
We focus exclusively on the CPA, as the results presented in Section~\ref{sec:classic} and Section~\ref{sec:direct} clearly show that it significantly outperforms the CILP in terms of scalability, particularly for larger instances. The evaluation focuses on instances with $n \geq 12$. This choice is motivated by the findings in \cite{ciacco2026CAF}, which indicate that quantum formulations based on the CILP achieve optimal solutions only for small instances ($n < 12$), while performance rapidly degrades for larger problem sizes, leading to non-negligible optimality gaps.
In this context, our objective is to advance the current state of the art by demonstrating that the proposed approach enables quantum methods to effectively address larger instances.
Within each iteration, a time limit of 5 seconds is imposed on the solver.
Table~\ref{tab:hybrid} reports the performance of the hybrid quantum–classical approach based on the CPA with CAF preprocessing. 

The table summarizes the computational results for TSP instances with an increasing number of vertices $V$, providing a comparison between the CILP and CPA approaches.
The table reports: 
\textbf{OF$_{\bm{avg}}$}: average objective value obtained by the hybrid quantum solver, 
\textbf{OF$_{\bm{dev}}$}: standard deviation of the objective value across runs, 
\textbf{Time$_{\bm{avg}}$}: average total computational time (in seconds), computed as the sum of the computational times over all iterations within each run, 
\textbf{Time$_{\bm{dev}}$}: standard deviation of the total computational time, 
\textbf{Solve$_{\bm{avg}}$}: average solver time (in seconds), computed as the sum of the solver times over all iterations within each run, 
\textbf{Solve$_{\bm{dev}}$}: standard deviation of the solver time, 
\textbf{Iters$_{\bm{avg}}$}: average number of iterations of the subtour elimination procedure, 
\textbf{Iters$_{\bm{dev}}$}: standard deviation of the number of iterations, 
\textbf{Cuts$_{\bm{avg}}$}: average number of generated subtour elimination constraints, 
\textbf{Cuts$_{\bm{dev}}$}: standard deviation of the number of constraints, 
\textbf{GAP}: percentage optimality gap with respect to the best solution found by the solver { with respect to the optimal solution reported in Table~\ref{tab:milp_results}}.
\color{black}
\begin{table}[h]
\centering
\resizebox{\columnwidth}{!}{
\begin{tabular}{c|cc cccc cc cc |c}
\toprule
\multirow{2}{*}{\textbf{V}} 
& \multicolumn{11}{c}{\textbf{CPA (With CAF)}} \\

\cmidrule(lr){2-12}
& 
 OF$_{avg}$ & OF$_{dev}$ 
& Time$_{avg}$ & Time$_{dev}$
& Solve$_{avg}$ & Solve$_{dev}$
& Iters$_{avg}$ & Iters$_{dev}$
& Cuts$_{avg}$ & Cuts$_{dev}$ & \textbf{GAP (\%)}\\

\midrule
12 &  4056,68 & 0,00  & 44,27 & 2,94  & 20,00 & 0,30  & 4  & 0 & 12 & 1 & 0\% \\
13 &  4564,46 & 0,00  & 43,64 & 4,50  & 19,96 & 0,25  & 4  & 0 & 13 & 0 & 0\% \\
14 &  4965,33 & 0,00  & 63,24 & 3,39  & 29,98 & 0,24  & 6  & 0 & 19 & 0 & 0\% \\
15 &  4967,30 & 0,00  & 54,60 & 3,86  & 25,10 & 0,27  & 5  & 0 & 19 & 0 & 0\% \\
20 &  5270,86 & 0,00  & 68,88 & 35,11 & 25,67 & 0,42  & 5  & 0 & 22 & 0 & 0\% \\
25 & 5460,94 & 0,00  & 114,46 & 14,41 & 48,42 & 3,76  & 9  & 1 & 35 & 2 & 0\% \\
30 &  6206,03 & 57,75 & 140,22 & 42,76 & 59,03 & 10,68 & 11 & 2 & 45 & 4 & 1\% \\
\bottomrule
\end{tabular}
}
\caption{ Computational results for the hybrid quantum--classical approach based on the CPA combined with CAF preprocessing.  
}
\label{tab:hybrid}
\end{table}

The results show that the proposed approach is able to reach optimal solutions for all instances up to $n = 25$, with a zero optimality gap and no variability across runs (standard deviation equal to zero). This indicates a high level of robustness and consistency of the method on small and medium-sized instances.
For the largest instance ($n = 30$), a slight degradation in performance is observed. In particular, the average objective value is 6206,03 compared to the optimal value of 6146,65, resulting in a very small optimality gap of 1\%. Thus, the gap remains extremely limited, confirming the effectiveness of the approach even at larger scales.
The total computational time, computed as the sum of the computational times over all iterations within each run, increases with the problem size, as expected. In particular, the average runtime grows from approximately 44 seconds for $n = 12$ to about 140 seconds for $n = 30$. A similar trend is observed for the solver time, which, being computed as the sum over all iterations within each run, increases from about 20 seconds to 59 seconds.
Moreover, the variability in computational time becomes more pronounced for larger instances. This is reflected in the increasing standard deviation, especially for $n \geq 20$, where the algorithm exhibits a less predictable runtime behavior. This effect is likely due to the increased size of the search space and the stochastic components of the hybrid solver.
The number of iterations required by the subtour elimination procedure grows moderately with the instance size. Specifically, it increases from an average of 4 iterations for $n = 12, 13$ to 11 iterations for $n = 30$. This trend is consistent with the theoretical bound proportional to $n/2$, confirming that the imposed iteration limit is appropriate and not restrictive in practice.
Similarly, the number of generated subtour elimination constraints increases steadily with $n$, from 12 constraints for $n = 12$ up to 45 for $n = 30$. This behavior reflects the progressive refinement of the solution through the addition of violated constraints. The relatively low number of cuts, compared to the exponential number required in the CILP, further highlights the effectiveness of the CPA.

Overall, the results demonstrate that the combination of CPA and CAF preprocessing enables the hybrid quantum–classical approach to achieve high-quality solutions with limited computational effort. The method scales effectively up to medium-large instances, maintaining optimality up to $n = 25$ and exhibiting only marginal degradation for $n = 30$. These findings confirm that the proposed framework successfully mitigates the complexity of the TSP, making it suitable for integration with quantum and hybrid optimization solvers.

\section{Conclusions}\label{sec:conclusions} 
In this work, we investigated the TSP within a framework combining classical optimization and QA approaches. The main challenge addressed concerns the exponential growth of subtour elimination constraints, which severely limits the scalability of both classical and quantum formulations. 
To overcome this limitation, we proposed an approach based on two complementary strategies: an iterative subtour elimination procedure (CPA) and a preprocessing phase (CAF). The CPA dynamically generates only the necessary subtour constraints, while CAF reduces the number of candidate arcs, leading to a more compact optimization model.
The computational results clearly demonstrate the effectiveness of the proposed framework. From a classical optimization perspective, the CPA drastically improves scalability, enabling the solution of instances that are otherwise intractable for the CILP. The integration of CAF further enhances performance by reducing the number of variables and computational time.
From a quantum optimization perspective, within direct QPU-based approaches, our results clearly demonstrate that the proposed CPA leads to a substantial improvement over the CILP, significantly enhancing both performance and robustness of quantum methods. In particular, it enables higher feasibility rates, better solution quality, and a more controlled growth of computational effort. This represents a significant improvement over standard QUBO-based approaches commonly adopted in the literature, where subtour elimination constraints are encoded a priori, leading to rapidly intractable models.
The hybrid quantum--classical approach demonstrates strong and consistent performance across all tested instances. In particular, the proposed framework achieves optimal solutions for medium-sized instances (up to $n = 25$) and maintains near-optimal performance for larger instances, with only marginal optimality gaps (around 1\%).These results clearly indicate that hybrid approaches currently represent the most effective and scalable strategy for leveraging quantum technologies in combinatorial optimization, significantly outperforming direct QPU-based methods and narrowing the gap with classical optimization techniques.

Future research directions include the investigation of more advanced encoding strategies for quantum annealers, the integration of adaptive parameter tuning mechanisms, and the exploration of decomposition techniques to further improve scalability. In addition, the evolution of quantum hardware is expected to progressively reduce current limitations, opening the way for more competitive pure quantum approaches.

\section*{Declaration of competing interest}
The authors declare that they have no known competing financial interests or personal relationships that could have appeared to
influence the work reported in this paper.

\section*{Data availability}
Data will be made available on request.


  \bibliography{bibliography}

@article{miller1960integer,
  title={Integer programming formulation of traveling salesman problems},
  author={Miller, Clair E. and Tucker, Albert W. and Zemlin, Richard A.},
  journal={Journal of the ACM},
  volume={7},
  number={4},
  pages={326--329},
  year={1960}
}

@book{applegate2006traveling,
  title={The Traveling Salesman Problem: A Computational Study},
  author={Applegate, David and Bixby, Robert and Chv{\'a}tal, Va{\v{s}}ek and Cook, William},
  publisher={Princeton University Press},
  year={2006}
}

@article{lin1973effective,
  title={An effective heuristic algorithm for the traveling-salesman problem},
  author={Lin, Shen and Kernighan, Brian W.},
  journal={Operations Research},
  volume={21},
  number={2},
  pages={498--516},
  year={1973}
}

@article{laporte1992tsp,
title = {The Traveling Salesman Problem: An Overview of Exact and Approximate Algorithms},
author = {Laporte, Gilbert},
journal = {European Journal of Operational Research},
volume = {59},
pages = {231--247},
year = {1992},
doi = {10.1016/0377-2217(92)90138-Y}
}

@incollection{cook2010integer,
  title={Fifty-plus years of combinatorial integer programming},
  author={Cook, William},
  booktitle={50 Years of Integer Programming 1958-2008: From the Early Years to the State-of-the-Art},
  pages={387--430},
  year={2009},
  publisher={Springer}
}

@article{ciacco2026CAF,
  title={Traveling Salesman Problem with a preprocessing method for classical and quantum optimization},
  author={Ciacco, Alessia and Pugliese, Luigi Di Puglia and Guerriero, Francesca},
  journal={arXiv preprint arXiv:2603.23290},
  year={2026}
}

@article{karg1964heuristic,
title = {A Heuristic Approach to Solving Travelling Salesman Problems},
author = {Karg, Robert and Thompson, G.},
journal = {Management Science},
volume = {10},
pages = {225--248},
year = {1964},
doi = {10.1287/MNSC.10.2.225}
}

@article{croes1958tsp,
title = {A Method for Solving Traveling-Salesman Problems},
author = {Croes, G. A.},
journal = {Operations Research},
volume = {6},
pages = {791--812},
year = {1958},
doi = {10.1287/OPRE.6.6.791}
}

@article{perez2024solving,
  author    = {P{\'e}rez Armas, Luis Fernando and Creemers, Stefan and Deleplanque, Samuel},
  title     = {Solving the resource constrained project scheduling problem with quantum annealing},
  journal   = {Scientific Reports},
  year      = {2024},
  volume    = {14},
  number    = {1},
  pages     = {16784},
  doi       = {10.1038/s41598-024-67168-6},
  url       = {https://doi.org/10.1038/s41598-024-67168-6}
}

@article{holliday2025advanced,
  title={Advanced Quantum Annealing Approach to Vehicle Routing Problems with Time Windows},
  author={Holliday, James B and Blount, Darren and Osaba, Eneko and Luu, Khoa},
  journal={arXiv preprint arXiv:2503.24285},
  year={2025}
}

@inproceedings{de2022hybrid,
  title={Hybrid quantum-classical heuristic for the bin packing problem},
  author={de Andoin, Mikel Garcia and Osaba, Eneko and Oregi, Izaskun and Villar-Rodriguez, Esther and Sanz, Mikel},
  booktitle={Proceedings of the Genetic and Evolutionary Computation Conference Companion},
  pages={2214--2222},
  year={2022}
}

@article{garcia2022comparative,
  title={Comparative Benchmark of a Quantum Algorithm for the Bin Packing Problem},
  author={Garcia-de-Andoin, Mikel and Oregi, Izaskun and Villar-Rodriguez, Esther and Osaba, Eneko and Sanz, Mikel},
  journal={arXiv preprint arXiv:2207.07460},
  year={2022}
}

@inproceedings{malviya2023logistics,
  title={Logistics network optimization using quantum annealing},
  author={Malviya, Gajendra and AkashNarayanan, B and Seshadri, Janani},
  booktitle={International Conference on Emerging Trends and Technologies on Intelligent Systems},
  pages={401--413},
  year={2023},
  organization={Springer}
}

@article{baldazzi2025variationaltsp,
  title={Resource-efficient variational quantum solver for the travelling salesman problem and its silicon photonics implementation},
  author={Baldazzi, Alessio and Azzini, Stefano and Pavesi, Lorenzo},
  journal={arXiv preprint arXiv:2511.02696},
  year={2025}
}

@article{qian2023qaoatsp,
title = {Comparative Study of Variations in Quantum Approximate Optimization Algorithms for the Traveling Salesman Problem},
author = {Qian, W. and Basili, R. and Eshaghian-Wilner, M. and Khokhar, A. and Luecke, G. and Vary, J.},
journal = {Entropy},
volume = {25},
year = {2023},
doi = {10.3390/e25081238}
}

@article{bell2025qaoarepresentations,
  title={A Comparison of Quadratic and Higher-Order Representations for QAOA},
  author={Bell, Kristina and Lowe, Adam and Coles, Emily and Ridgway, Nathanael and Marshall, Gillian},
  journal={arXiv preprint arXiv:2509.20127},
  year={2025}
}

@inproceedings{venturelli2016job,
  title={Job shop scheduling solver based on quantum annealing},
  author={Venturelli, Davide and Marchand, D and Rojo, Galo},
  booktitle={Proc. of ICAPS-16 Workshop on Constraint Satisfaction Techniques for Planning and Scheduling (COPLAS)},
  pages={25--34},
  year={2016}
}

@article{ciacco2026facility,
  title={Quantum annealing for the two-level facility location problem},
  author={Ciacco, Alessia and Guerriero, Francesca and Saccomanno, Francesco Paolo},
  journal={Future Generation Computer Systems},
  volume={174},
  pages={107961},
  year={2026},
  publisher={Elsevier}
}

@inproceedings{osaba2025quantum,
  title={Quantum-Assisted Automatic Path-Planning for Robotic Quality Inspection in Industry 4.0},
  author={Osaba, Eneko and Garrote, Estibaliz and Miranda-Rodriguez, Pablo and Ciacco, Alessia and Cabanes, Itziar and Mancisidor, Aitziber},
  booktitle={2025 IEEE International Conference on Quantum Computing and Engineering (QCE)},
  volume={2},
  pages={388--389},
  year={2025},
  organization={IEEE}
}

@inproceedings{ciacco2025steiner,
  title={Steiner Traveling Salesman Problem with Quantum Annealing},
  author={Ciacco, Alessia and Guerriero, Francesca and Osaba, Eneko},
  booktitle={Proceedings of the Genetic and Evolutionary Computation Conference Companion},
  pages={2412--2418},
  year={2025}
}

@article{jain2021tsp,
  author  = {Jain, Siddharth},
  title   = {Solving the Traveling Salesman Problem on the D-Wave Quantum Computer},
  journal = {Frontiers in Physics},
  volume  = {9},
  pages   = {760783},
  year    = {2021},
  doi     = {10.3389/fphy.2021.760783}
}

@article{warren2020tsp,
  author  = {Warren, Richard H.},
  title   = {Solving Combinatorial Problems by Two D-Wave Hybrid Solvers: A Case Study of Traveling Salesman Problems in the TSP Library},
  journal = {arXiv preprint arXiv:2106.05948},
  year    = {2021},
  url     = {https://arxiv.org/abs/2106.05948}
}

@article{bochkarev2026quantum,
  author  = {Bochkarev, Alexey and Heese, Raoul and J{\"a}ger, Sven and Schiewe, Philine and Sch{\"o}bel, Anita},
  title   = {Quantum Computing for Discrete Optimization: A Highlight of Three Technologies},
  journal = {European Journal of Operational Research},
  volume  = {329},
  pages   = {747--766},
  year    = {2026},
  doi     = {10.1016/j.ejor.2025.07.063}
}

@incollection{karp1972reducibility,
  title={Reducibility among combinatorial problems},
  author={Karp, Richard M},
  booktitle={50 Years of Integer Programming 1958-2008: from the Early Years to the State-of-the-Art},
  pages={219--241},
  year={2009},
  publisher={Springer}
}

@article{reinelt1991tsplib,
  title={TSPLIB—A traveling salesman problem library},
  author={Reinelt, Gerhard},
  journal={ORSA journal on computing},
  volume={3},
  number={4},
  pages={376--384},
  year={1991},
  publisher={Informs}
}

@misc{dwave_cqm_solver,
  author       = {{D-Wave Systems}},
  title        = {New Hybrid Solver: Constrained Quadratic Model},
  year         = {2021},
  url          = {https://support.dwavesys.com/hc/en-us/articles/4410049190807-New-Hybrid-Solver-Constrained-Quadratic-Model},
  note         = {Accessed: 2026-03-06}
}

@misc{dwave_cqm_parameters,
  author       = {{D-Wave Systems}},
  title        = {CQM Solver Parameters},
  year         = {2024},
  url          = {https://docs.dwavequantum.com/en/latest/industrial_optimization/solver_cqm_parameters.html},
  note         = {D-Wave Documentation, Accessed: 2026-03-06}
}

@article{ciacco2025steinerv2,
  title={Steiner Traveling Salesman Problem with Time Windows and Pickup-Delivery: integrating classical and quantum optimization},
  author={Ciacco, Alessia and Guerriero, Francesca and Osaba, Eneko},
  journal={arXiv preprint arXiv:2508.17896},
  year={2025}
}

@article{ciacco2025Educational,
  title={Quantum Annealing for Staff Scheduling in Educational Environments},
  author={Ciacco, Alessia and Guerriero, Francesca and Osaba, Eneko},
  journal={arXiv preprint arXiv:2510.12278},
  year={2025}
}

@article{smith2025travelling,
  title={The travelling salesperson problem and the challenges of near-term quantum advantage},
  author={Smith-Miles, Kate A and Hoos, Holger H and Wang, Hao and B{\"a}ck, Thomas and Osborne, Tobias J},
  journal={Quantum Science and Technology},
  volume={10},
  number={3},
  pages={033001},
  year={2025},
  publisher={IOP Publishing}
}

@article{willsch2022benchmarking,
  title={Benchmarking Advantage and D-Wave 2000Q quantum annealers with exact cover problems},
  author={Willsch, Dennis and Willsch, Madita and Gonzalez Calaza, Carlos D and Jin, Fengping and De Raedt, Hans and Svensson, Marika and Michielsen, Kristel},
  journal={Quantum Information Processing},
  volume={21},
  number={4},
  pages={141},
  year={2022},
  publisher={Springer}
}

@inproceedings{bertuzzi2024evaluation,
  title={Evaluation of quantum and hybrid solvers for combinatorial optimization},
  author={Bertuzzi, Amedeo and Ferrari, Davide and Manzalini, Antonio and Amoretti, Michele},
  booktitle={Proceedings of the 21st ACM International Conference on Computing Frontiers},
  pages={232--239},
  year={2024}
}

@article{osaba2025d,
  title={D-wave’s nonlinear-program hybrid solver: Description and performance analysis},
  author={Osaba, Eneko and Miranda-Rodriguez, Pablo},
  journal={IEEE Access},
  year={2025},
  publisher={IEEE}
}

@article{dantzig1954solution,
  title={Solution of a large-scale traveling-salesman problem},
  author={Dantzig, George and Fulkerson, Ray and Johnson, Selmer},
  journal={Journal of the operations research society of America},
  volume={2},
  number={4},
  pages={393--410},
  year={1954},
  publisher={INFORMS}
}

@article{abbas2024challenges,
  title={Challenges and opportunities in quantum optimization},
  author={Abbas, Amira and Ambainis, Andris and Augustino, Brandon and B{\"a}rtschi, Andreas and Buhrman, Harry and Coffrin, Carleton and Cortiana, Giorgio and Dunjko, Vedran and Egger, Daniel J and Elmegreen, Bruce G and others},
  journal={Nature Reviews Physics},
  pages={1--18},
  year={2024},
  publisher={Nature Publishing Group}
}

@article{ciacco2025review,
  title={Review of quantum algorithms for medicine, finance and logistics},
  author={Ciacco, Alessia and Guerriero, Francesca and Macrina, Giusy},
  journal={Soft Computing},
  volume={29},
  number={4},
  pages={2129--2170},
  year={2025},
  publisher={Springer}
}

@manual{gurobi-doc,
  title = {Gurobi Optimizer Reference Manual},
  author = {{Gurobi Optimization, LLC}},
  year = {2024},
  url = {https://www.gurobi.com/documentation/}
}

\end{document}